\documentclass{osa-article}

\usepackage{siunitx}
\usepackage{subfigure}
\usepackage[title]{appendix}

\newcommand{\figref}[1]{Fig.~\ref{#1}}
\renewcommand{\eqref}[1]{Eq.~(\ref{#1})}

\newcommand{\geff}{G_\mathrm{eff}}
\newcommand{\PSIin}{\psi_\mathrm{in}}
\newcommand{\PSImid}{\psi_\mathrm{mid}}
\newcommand{\PSIout}{\psi_\mathrm{out}}

\journal{oe}
\articletype{Research Article}
\begin{document}

\title{Microscopy with heralded Fock states}

\author{Maria Gieysztor,\authormark{1} Joshua Nepinak,\authormark{2} \mbox{Christopher J. Pugh},\authormark{1,2,3} and Piotr Kolenderski\authormark{1, *}}

\address{\authormark{1}Institute of Physics, Faculty of Physics, Astronomy and Informatics, Nicolaus Copernicus University in Toru\'{n}, Grudzi\k{a}dzka 5, 87-100 Toru\'{n}, Poland\\
\authormark{2}Department of Physics and Astronomy, Brandon University, 270-18 Street, Brandon, MB, R7A 6A9, Canada\\
\authormark{3}Newman Theological College, 10012-84 Street NW, Edmonton, AB, T6A 0B2, Canada}

\email{\authormark{*}kolenderski@umk.pl} %% email address is required



\begin{abstract}
	We consider a microscopy setting where quantum light is used for illumination. Spontaneous parametric down conversion (SPDC) is used as a source of a heralded single photon, which is quantum light prepared in a Fock state. We present  analytical formulas for the spatial mode tracking along with the heralded and non-heralded mode widths. The obtained analytical results are supported by numerical calculations and the following discussion taking into account realistic setup parameters such as finite-size optics and finite-size single-photon detectors. This allows us to observe that the diffraction limit can be approached with simultaneous alleviation of the photon loss leading to increased signal-to-noise ratio -- a factor limiting practical applications of quantum light. Additionally, it is shown that the spatial resolution can be manipulated by carefully preparing  the amplitude and phase of the spatial mode profile of the single photon at the input to the microscope objective. Here, the spatial entanglement of the biphoton wavefunction or adaptive optics can be applied for spatial mode shaping. Analytical dependencies between the incident and focused spatial mode profiles parameters are provided. 
\end{abstract}

%%%%%%%%%%%%%%%%%%%%%%%%%%  body  %%%%%%%%%%%%%%%%%%%%%%%%%%
\section{Introduction}

Complete control of a light beam consists of the ability to  manipulate all its degrees of freedom, comprising the wavelength, polarization, intensity, and the spatial mode profile (SMP). Wavelength and polarization are especially important for investigating light-matter interaction as they have to match the absorber's energy structure and dipole moment orientation, respectively. Additional information can, in principle, be extracted from a sample by using quantum light illumination with well-defined photon number statistics. The process of SPDC is a well-known method for generating photons in a Fock state. In addition, the SMP of a light beam directly determines the lateral resolution of a microscope, which is bound by finite-size optics and results in the fundamental diffraction limit. It is given by \textit{eg.}~the Rayleigh criterion~\cite{Hecht2002, Saleh2007}, $0.61 \lambda /$NA, where $\lambda$ is the wavelength of the illuminating beam and NA stands for the numerical aperture of the optical setup, typically the microscope objective (MO). For the visible range and typical optics, it gives a spot size of  the order of \mbox{$200$ nm}. Such a characterization of the light beam is primarily interesting from the fundamental point of view, but additionally, it is important in the context of quantum microscopy, where the classical resolution limit can be beaten when exploiting NOON states \cite{Simon2016, Boto2000}.

Quantum ghost imaging \cite{Klyshko1988, Pittman1995, Scarcelli2008} is a method allowing one to control the light interacting with an atomic system in a microscopy setting. There are a few technical challenges to overcome to make it efficient in a real experimental setup. High-brightness photon-pair sources are one of the elements \cite{Meyer-Scott2018}. Moreover, the photon loss in the experimental setup is critical for experiments relying on coincidence detection measurements, as the inefficiency scales quadratically with attenuation. However, in a classic approach achieving the diffraction limit requires the lateral spatial mode of the light entering a microscope objective to be comparable with the size of its entrance aperture. This in turn introduces photon loss, which reduces the signal-to-noise ratio of the final measurement.

In this work, we focus on the resolution and the photon loss problem of a quantum ghost imaging setup, where the sample is illuminated by SPDC-generated heralded single photons. In contrast to the literature of the subject \cite{Moreau2018, Schneeloch2016a},  we investigate not only the fundamental limits of the resolution, but we also analyze the illuminating beam SMP parameters together with the photon loss problem. The article is organized as follows. In the first part, we present an introductory discussion regarding the impact of the phase profile of a light beam entering a microscope objective on the resulting focused spot size. Next, in the main part of the work, we perform a systematic theoretical analysis of the single photon SMP in the framework of quantum ghost imaging introduced by Pittmann \textit{et al.}~\cite{Pittman1995}. Using mathematical tools developed by Abouraddy \textit{et al.}~\cite{Abouraddy2002}, a general Gaussian form of a correlated biphoton wavefunction, describing a pair of photons produced in the SPDC process, is propagated through an optical setup. The signal photon SMP is studied in the heralding scheme and a comparison with the non-heralding scheme is provided. Our conclusions are supported by numerical calculations and analytical formulas. The calculations were made for realistic experimental parameters.

\section{Single mode propagation}
\label{sec:single_mode_propagation}

Let us assume a Gaussian beam, which is described by a function, $\psi_1(x)$, in the following form:
\begin{equation}
	\psi_1(x) = N_1 \exp{\left(- \frac{(x-x_1)^2}{\sigma_1^2} + i  \phi(x) \right)},
	\label{eq:single_beam_profile}
\end{equation}
where $\sigma_1$ is the spatial mode width, $x_1$ is its central position, $N_1$ stands for a normalization constant, and 
\begin{equation}
	\phi(x) = A x^2 + B x + C
	\label{eq:phase}
\end{equation}
is the phase. The free-space propagator that we apply here is of the form $ \exp\left( -{i \pi (x' - x)^2}/{\lambda d}\right)$ \cite{Saleh2007}, where $d$ is the propagation distance and $\lambda$ is the wavelength. Similarly, the respective transformation of a lens of focal length $f$ reads $\exp\left( {i \pi x^2}/{\lambda f}\right)$. We consider propagation through a lens of infinite aperture. The resulting amplitude at a distance equal to the lens focal length, $f$, is then described by:
\begin{equation}
	\psi_2(x)= N_2 \exp{(-\frac{(x - x_2)^2}{\sigma_2^2}+ i  \phi_2(x))},
\end{equation}
where the output beam width equals to:
\begin{equation}
	\sigma_2 = \frac{f \lambda}{\pi \sigma_1} \sqrt{1 + A^2 \sigma_1^4},
	\label{eq:sigma2}
\end{equation}
and its central position reads: 
\begin{equation}
	x_2 = -\frac{f \lambda}{2 \pi}(2 A x_1 + B).
	\label{eq:x2}
\end{equation}
The normalization constant $N_2$ and the resulting phase $\phi_2(x)$ are not important for  further discussion. One can clearly see that $\sigma_2$ depends not only on the width of the SMP entering the lens, $\sigma_1$, but also on the quadratic phase coefficient, $A$, of the incoming light. As a result, the quadratic phase makes the resulting beam wider. The central position of the propagated beam, however, depends not only on the central position of the incoming beam $x_1$, but also on the quadratic, $A$, and linear, $B$, phase coefficients. This leads to the conclusion that by imprinting an appropriate input beam phase both the width and the central position of a Gaussian beam propagated through a lens can be manipulated. In particular, a phase-altering device like a spatial light modulator (SLM) could be used for beam manipulation. Although beam steering with SLMs is a known method \cite{Engstrom2008}, here we provide an analytical description of how the SMP after the lens depends on the phase of the incident light beam. Please note that the discussion presented above can apply to both classical light and single photons.

\section{Biphoton propagation}
\label{sec:biphoton_propagation}

\begin{figure}[htb]
	\centering
	\includegraphics[width=0.55\linewidth]{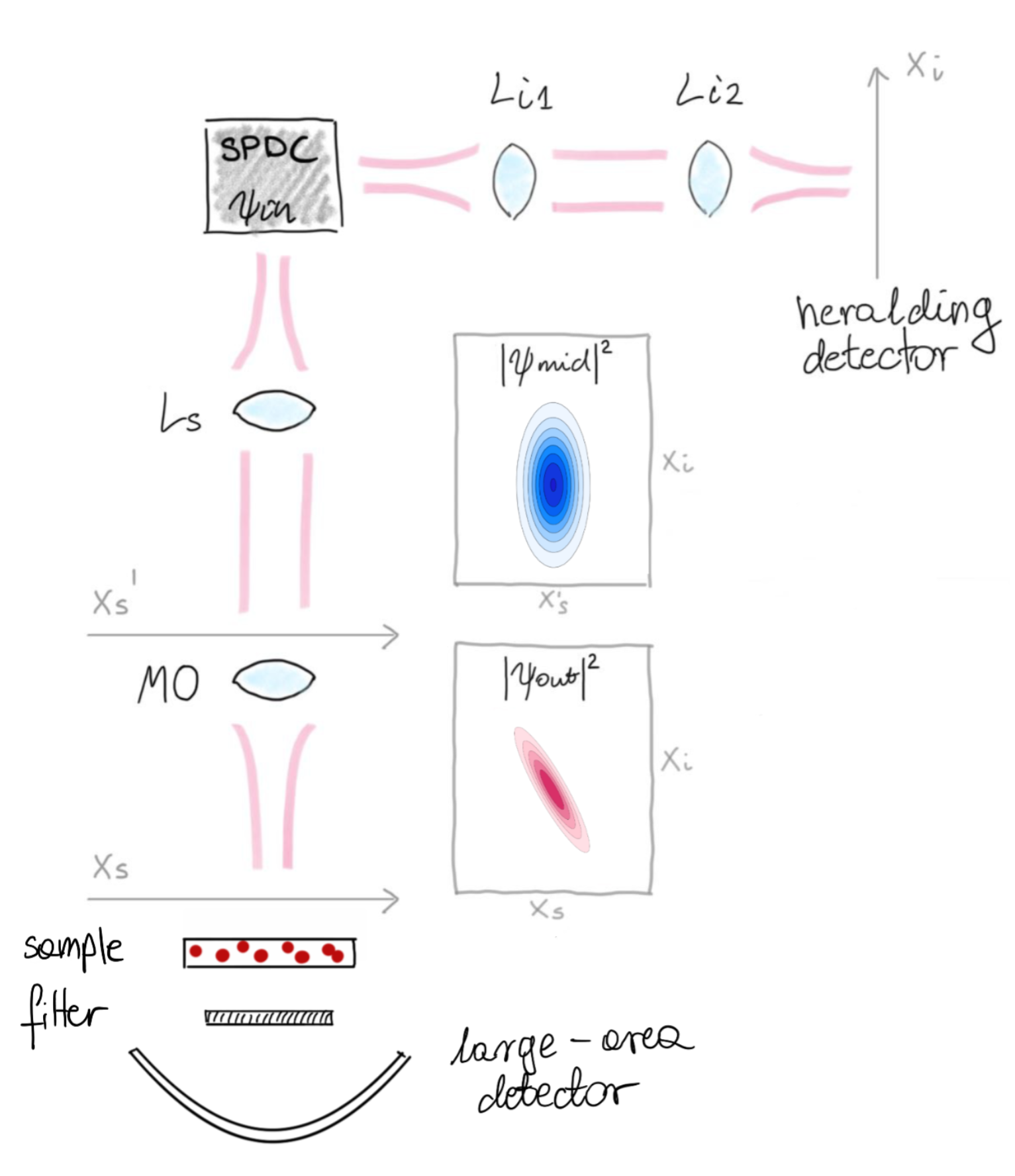}
	\caption{\textbf{Experimental setup concept.} The input biphoton wavefunction, $\PSIin$, is generated in a nonlinear crystal by means of SPDC process. The signal photon propagates through free-space ($d_{s1}$), signal arm lens $L_\mathrm{s}$ ($f_{s}$), free-space ($d_{s2}$), microscope objective MO ($f_{MO}$), again free-space ($d_{s3}$) and hits a sample. The fluorescence from the sample can be collected by a large-area detector. The idler photon propagates through free-space ($d_{i1}$), first idler arm lens $L_\mathrm{i1}$ ($f_{i1}$), free-space ($d_{i1}$), second idler arm lens $L_{i2}$ ($f_{i2}$), again free-space ($d_{i3}$) and hits a detector. The wavefunction propagated through free-space, first lens, and again free space on the signal arm and through the whole idler arm is denoted by $\PSImid(x_\mathrm{s}', x_\mathrm{i})$, whereas the output wavefunction propagated through the whole signal and idler arms is denoted by $\PSIout(x_\mathrm{s}, x_\mathrm{i})$. Typical joint probability distributions $|\PSImid(x_\mathrm{s}',x_\mathrm{i})|^2$ (blue) and $|\PSIout(x_\mathrm{s}, x_\mathrm{i})|^2$ (pink) computed for a correlation of $\rho = 0.9$ are presented in the contour plots.}
	\label{Fig:Experimental-setup-concept}
\end{figure}

Here we present the framework for the propagation of a pair of photons in a microscope setup given in \figref{Fig:Experimental-setup-concept}. We assume that the probability density amplitude of generating a pair in a nonlinear crystal, where the idler and signal directions are given by $\kappa_\mathrm{i}, \kappa_\mathrm{s}$, is of a general Gaussian form~\cite{Schneeloch2016a, Pugh2016} and reads:
\begin{equation}
	\tilde{\PSIin} (\kappa_\mathrm{s}, \kappa_\mathrm{i}) = \mathcal{N} \exp\left({- \frac{1}{4}\left(\frac{\kappa_\mathrm{s}^2}{\delta_\mathrm{s}^2} + \frac{\kappa_\mathrm{i}^2}{\delta_\mathrm{i}^2} - \frac{2 \kappa_\mathrm{i}\kappa_\mathrm{s} \rho}{\delta_\mathrm{i} \delta_\mathrm{s}}\right)}\right),
	\label{eq:wavefunction2}
\end{equation}
where $\delta_\mathrm{s}$, $\delta_\mathrm{i}$ stand for the mode profile radii of the signal and idler photons, $\rho$ is the Pearson coefficient, which is a measure of quantum correlation (entanglement), and $\mathcal{N}$ is the normalization constant. The relationship between \eqref{eq:wavefunction2} and the double-Gaussian approximation of the sinc-Gaussian biphoton wavefunction %that is commonly used in this context, can be found by a direct calculation.
can be found in Refs~\cite{Zhong2016, Schneeloch2016a}. The parametrization used in the present paper however, enables direct access to the correlation parameter, $\rho$, which impacts the SMP of the signal photon that we wish to investigate. Nevertheless, it leads to correlation-dependent marginal distributions widths $\delta_s/\sqrt{1-\rho^2}$ and $\delta_i/\sqrt{1-\rho^2}$ for signal and idler photons, respectively. Please note that the formula given in \eqref{eq:wavefunction2} and the following discussion concern a single transverse spatial dimension for each of the photons, the signal, and the idler.

The pair generated in a nonlinear crystal propagates through the optical setup depicted in Fig.~\ref{Fig:Experimental-setup-concept}. The idler photon, used as a reference, travels through a system of two lenses and is finally measured  by a single-photon detector, which heralds the signal photon. The distances between the source, the two lenses ($L_{s1}, L_{s2}$) and the heralding plane in the idler photon path are $d_\mathrm{i1}$,  $d_\mathrm{i2}$, and $d_\mathrm{i3}$, respectively. The signal photon, which is used as a probe, propagates through a lens ($L_s$) and later is focused on a sample by a MO. Changing the focal length of the $L_s$ lens, $f_s$, allows for simple manipulation of the signal photon spatial mode before the MO. The fluorescence from the sample can be spectrally filtered and registered by a large-area single-photon detector, providing no spatial resolution. The distances between the source, first lens, MO, and the sample in the signal photon path are $d_\mathrm{s1}$,  $d_\mathrm{s2}$ and $d_\mathrm{s3}$, respectively. The focal lengths of the lenses in the idler arm are $f_\mathrm{i1}$ and $f_\mathrm{i2}$ and in the signal arm are $f_\mathrm{s}$ and $f_\mathrm{MO}$. The formulas for the free-space propagator and the respective lens propagator were already given in the previous section. All of the lenses are assumed to be thin and their apertures are $R_\mathrm{i1}, R_\mathrm{i2}$ for the idler and $R_\mathrm{s}, R_\mathrm{MO}$ for the signal arm. All of the optical system  parameters listed so far are combined into a parameter vector $\vec{p}$ for convenience of notation. This allows us to symbolically represent the resulting wavefunction of the idler photon in the heralding detector plane and the signal photon at the sample plane in the following compact form \cite{Saleh2007, Sedziak17}:
\begin{equation}
	\PSIout (x_\mathrm{s}, x_\mathrm{i}) =\\
	\int_{-\infty}^{\infty} dx_\mathrm{s}' \int_{-\infty}^{\infty} dx_\mathrm{i}' \; \geff(x_\mathrm{s}', x_\mathrm{i}', x_\mathrm{s}, x_\mathrm{i}, \vec{p}) \;  \PSIin(x_\mathrm{s}',x_\mathrm{i}') .
	\label{eq:wavefinction-after-propagation-general-idea}
\end{equation}
Here, $\PSIin(x_\mathrm{s}',x_\mathrm{i}')$ stands for the Fourier transform of the initial biphoton wavefunction given in \eqref{eq:wavefunction2} and $\geff(x_\mathrm{s}', x_\mathrm{i}', x_\mathrm{s}, x_\mathrm{i}, \vec{p})$ stands for the effective transformation including respective free-space propagation and lens transformations in both arms.

The biphoton wavefunction, as presented above, allows us to analyze the influence of the spatial entanglement on the performance of the single photon illumination. When the idler photon is measured in the heralding plane at position $x_\mathrm{i}$ by a point-like detector (PLD), the probability density of the signal photon detection at the sample plane at position $x_s$ is given by:
\begin{equation}
	p_\mathrm{PLD} (x_\mathrm{s}; x_\mathrm{i}) = |\PSIout(x_\mathrm{s}, x_\mathrm{i})|^2.
	\label{eq:p:pld}
\end{equation}
The probability to detect the signal photon heralded by a finite size detector (FSD) can be evaluated  by integrating the probability density, given by the formula above, over the active area of the heralding device. In turn, if the information about the idler photon's location is disregarded, the probability density, $p_s(x_\mathrm{s})$, of detecting the signal photon at position $x_s$ can be found by tracing over the idler photon's position, which results in the following formula:
\begin{equation}
	p_s(x_\mathrm{s}) = \int_{-\infty}^{\infty} dx_\mathrm{i} |\PSIout(x_\mathrm{s}, x_\mathrm{i})|^2.
	\label{eq:p:nh}
\end{equation}
The spatial mode intensity profile of the idler photon, $p_i (x_\mathrm{i})$, can be computed analogically. The two formulas, \eqref{eq:p:pld} and \eqref{eq:p:nh}, constitute the framework for our analysis of the effects of entanglement and spatially resolved detection, which can be used to control the illumination. \eqref{eq:p:pld} is related to the heralding and  \eqref{eq:p:nh} to the non-heralding scenario.

\section{Results}

Here we perform the analysis of the signal photon SMP in the regime of experimentally accessible setup parameters (see Sec. \ref{Sec:quantitative_analysis} for details). Due to the finite lateral size of lenses, an accurate analysis requires numerical calculations. Nevertheless, to get a better understanding of the presented problem, a simplified analysis assuming infinite lenses will be performed first, as in this case the calculations yield analytical formulas. Such assumption can be justified, as for the considered ranges of parameters, the mode profiles entering the signal lens and both idler arm lenses are much smaller than typical lens apertures. This argument however, does not apply to the case of the MO entrance pupil diameter, which can be comparable or even smaller, than the SMP. Even though such simplification certainly introduces substantial limitations, like the absence of the diffraction limit, a great part of the outcomes still remain valid in the finite lens scenario. In addition, it enables an in-depth analysis of the considered problem. After the simplified, qualitative analysis, the more accurate, quantitative analysis, will be presented.

\subsection{Qualitative analysis}
\label{subsec:Qualitative_analysis}

The signal photon SMP will be analyzed in two characteristic spots: just before the MO and in the focal plane of the MO, which is the sample plane. This can be done by evaluating analytically \eqref{eq:wavefinction-after-propagation-general-idea}, which is possible due to the Gaussian form of all the involved functions. The lenses were placed at the respective focal distances, which corresponded to $d_\mathrm{s1} = f_\mathrm{s}$, $d_\mathrm{s3} = f_\mathrm{MO}$, $d_\mathrm{i1} = f_\mathrm{i1}$ and $d_\mathrm{i3} = f_{i3}$.

First, we analyze the SMP of the signal photon at the entrance to the MO. This is done by  using in \eqref{eq:wavefinction-after-propagation-general-idea} a propagator related to the full path propagation in the idler arm and propagation up to the MO in the signal arm. Then the width of the non-heralded signal photon SMP before the MO reads:
\begin{equation}
	\frac{1}{\pi }{\sqrt{\frac{\pi ^2 (d_{s2}-f_{s})^2}{\delta_s^2 f_{s}^2}+\frac{\delta_s^2 f_{s}^2 \lambda^2}{1-\rho^2}}},
	\label{eq:width_before_MO_non_heralded}
\end{equation}
whereas the width of the signal photon SMP before the MO heralded by a PLD in the idler arm is given by: 
\begin{equation}
	\frac{1}{\pi }\sqrt{\frac{\pi^2(d_{s2}-f_s)^2}{\delta_s^2 f_s^2}(1-\rho^2)+\frac{ \delta_s^2f_s^2 \lambda^2}{ 1-\rho^2}}.
	\label{eq:width_before_MO_heralded}
\end{equation}
The derivation details of the above-given formulas can be found in Appendix~\ref{app:derivation}. Please note that the width of the SMP of the signal photon heralded by a PLD does not depend on the heralding detector position, $x_i$. Heralding at different $x_i$ however, has impact on the mode center location before the MO. In consequence, it is responsible further for the spatial mode central location after the full propagation as discussed in Sec. \ref{sec:single_mode_propagation}. The dependence of the non-heralded and heralded widths on the distance between the lenses in the signal arm is very weak for reasonable values of $d_{s2}$, reachable on a typical optical table. This corresponds to a width difference on a level below 1.5\% when changing the lens distance from $5$~cm to $2$~m. However, for a particular choice of $d_{s2} = f_{s}$ the formulas given in \eqref{eq:width_before_MO_non_heralded} and \eqref{eq:width_before_MO_heralded} both simplify to: ${\delta_s f_{s} \lambda}/{\pi \sqrt{1-\rho ^2}}$. This means that there is no correlation in the amplitude of the biphoton wavefunction. Hence, no narrowing nor change of the central position of the SMP can be observed.

The SMP widths of the non-heralded signal photon after full propagation at the sample plane can be computed as:
\begin{equation}
	\frac{1}{\delta_s}\frac{f_{MO}}{f_{s}}
	\label{eq:non-herladed-width-after-MO}
\end{equation}
in the non-heralded case and as
\begin{equation}
	\frac{1}{\delta_s}\frac{f_{MO}}{f_{s}}\sqrt{1-\rho ^2}
	\label{eq:herladed-width-after-MO}
\end{equation}
in the heralded case. The derivation details of the above-given formulas can be found in Appendix~\ref{app:derivation}. When comparing the widths obtained above it can be seen that the narrowing factor reads: $\sqrt{1-\rho^2}$. This is, as mentioned at the beginning of this section, only valid for an infinite lens scenario. In the case where finite lenses were used, the heralding ratio would be limited by the diffraction limit. It should be noted that these widths do not depend on the distance between the lenses $d_{s2}$ nor the heralding position $x_i$. What the heralding position influences is the signal photon SMP center location. This dependence  is given by the formula:
\begin{equation}
	x_\mathrm{sC} = - \frac{f_\mathrm{i1} f_\mathrm{MO}}{f_\mathrm{s} f_\mathrm{i2}} \frac{\delta_\mathrm{i}}{\delta_\mathrm{s}} \rho  \cdot x_\mathrm{i}.
	\label{eq:EAST}
\end{equation} 

As already mentioned, under typical experimental conditions, the widths of the SMPs of the heralded and non-heralded signal photons before the MO weakly depend on the distance $d_{s2}$. What, however, significantly depends on $d_{s2}$ is the phase of the heralded signal photon before the MO. This phase in turn impacts further the SMP of the signal photon at the sample plane as discussed in Sec. \ref{sec:single_mode_propagation}. To see that let us consider the wavefunction before the MO in more detail including not only its amplitude but also its phase profile.

The amplitude of the biphoton wavefunction before the MO is given by a two-dimensional Gaussian function of exactly the same structure as the input wavefunction given in \eqref{eq:wavefunction2}. The acquired phase however, is given by a general quadratic function of the position variables, $x_s$ and $x_i$, of the signal and idler photons, $a x_s^2 + b x_i x_s + c x_i^2$. When heralded by a PLD, $x_i=$const., the signal photon wavepacket before the MO becomes a pure state, where a general description like in Sec.~\ref{sec:single_mode_propagation} can be applied. Then the amplitude reduces to a Gaussian function like in \eqref{eq:single_beam_profile} and the phase to a quadratic function like in \eqref{eq:phase}. The formulas for the phase parameters, $A$, $B$, and $C$, with explicit dependence on the optical setup parameters, were derived and can be found below

\begin{equation}
	A = \frac{\pi ^3 \left(\rho ^2-1\right)^2 (f_s - d_{s2})}{\lambda X},
	\label{eq:A}
\end{equation}

\begin{equation}
	B = \frac{2 \pi  \delta_i \delta_s^3 f_{i1} f_s^3 \lambda  \rho }{f_{i2} X} * x_i,
	\label{eq:B}
\end{equation}

\begin{equation}
	C = \frac{\pi  \left(\delta_i^2 \delta_s^2 f_{i1}^2 f_s^2 \lambda ^2 \rho ^2 (f_s-d_{s2})/X+f_{i1}+f_{i2} -d_{i2}\right)}{f_{i2}^2 \lambda } * x_i^2,
	\label{eq:C}
\end{equation}

\noindent where

\begin{equation}
	X = \pi ^2 (\rho^2-1)^2 \left( (f_s - d_{s2})^2 + \frac{f_s^4 \delta_s^4 \lambda^2}{(\rho^2 - 1)^2 \pi^2} \right).
	\nonumber
\end{equation}
The derivation details can be found in Appendix \ref{app:derivation}. It can be easily seen that the distance between the $L_s$ lens and the MO, $d_{s2}$, influences both the $A$ and $B$ parameters. In particular, if this distance satisfies the condition $d_{s2} = f_{s}$, the $A$ parameter vanishes and the $B$ parameter takes its maximal value. Applying here the conclusions from Sec. \ref{sec:single_mode_propagation} one can clearly see, that the signal photon's SMP will become the smallest for $A=0$ (see \eqref{eq:sigma2}). Additionally, from  \eqref{eq:x2} one concludes that the location of the center of the heralded signal photon wavepacket depends not only on the central position of the signal photon spatial mode entering the MO but also on the quadratic, $A$, and linear, $B$, phase parameters. This means the condition $d_{s2} = f_{s}$ leads to the highest sensitivity of remote steering of the signal photons beam.  On the other hand, when fixing $d_{s2}\neq f_s$, the quadratic phase parameter, $A$, becomes nonzero. It results in the SMP of the signal photon at the sample plane to be broader than its minimal value obtained for $A=0$. This effect in principle can be reversed by introducing a phase-altering optical device which can modify the phase of a photon entering the MO. Moreover, tuning both the $A$ and $B$ parameters would enable wavepacket focusing on different positions in the detection plane. Although introducing an additional optical device comes with more photon loss, it might be considered for example as an alternative to standard scanning methods in standard microscopy.

The conclusions resulting from the analytical discussion presented above are entirely valid only when the incident photons SMPs are much smaller than the entrance pupil diameter of the lenses in the setup. It is crucial primarily when discussing the signal photon SMP of the signal photon after the full propagation. To show the impact of the spatial entanglement on the SMP of the signal photon in the sample plane exactly numerical simulations with finite-size optics and experimentally accessible parameters were necessary.

\subsection{Quantitative analysis}
\label{Sec:quantitative_analysis}

Let us now consider typical experimental parameters including finite-size lenses. The mode profiles of the photons generated by an example SPDC source were set to be equal, $\delta_\mathrm{s} = \delta_\mathrm{i} = {1}/{70}$ $\mu$m$^{-1}$. These values are typical for a $5$~mm thick BBO crystal pumped with a $400$ nm laser beam, focused to a spot of $100$~$\mu$m diameter. This leads to the Pearson coefficient of approximately $0.9$ \cite{Edgar2012,Schneeloch2016a}. Spectrally degenerate SPDC was considered and the wavelengths of the idler and signal photons were set to $800$~nm. The idler arm lenses' focal lengths were set to $f_\mathrm{i1} = 75$ mm and $f_\mathrm{i2} = 1000$ mm. They were chosen in such a way, that together with the choice of the SPDC source parameters, the SMP of the idler photon in the detection plane was big enough so that the detection in the heralding scenario by a commercially available detector, of an active area $100$~$\mu$m, could be considered as PLD. To vary the signal photon SMP before the MO, $\sigma_\mathrm{mid}$, simulation over a range of  $f_\mathrm{s} \in (100, 1000)$ mm was run. Such choice of the $L_s$ lens focal lengths allowed for the $\sigma_\mathrm{mid}$ to be  greater or comparable with the MO entrance aperture ($f_\mathrm{s} = 1000$~mm) or significantly smaller ($f_\mathrm{s} = 100$~mm).  All of the lenses had finite apertures of the  diameters: $2 \cdot R_\mathrm{s} = 2 \cdot R_\mathrm{i1} = 25$~mm and $2 \cdot R_\mathrm{i2} = 50$~mm. The parameters of the MO were set to \mbox{$f_\mathrm{MO} = 4$ mm} and $2 \cdot R_\mathrm{MO} = 5.2$~mm which leads to a typical NA. All of the lenses were placed at the respective focal distances, which corresponded to $d_\mathrm{s1} = f_\mathrm{s}$, $d_\mathrm{s3} = f_\mathrm{MO}$, $d_\mathrm{i1} = f_\mathrm{i1}$ and $d_\mathrm{i3} = f_{i3}$. The distances between the lenses were set to the same values for both arms, $d_\mathrm{s2} = d_\mathrm{i2} = 800$ mm. The simulation results turned out not to be sensitive to this parameter in a reasonable range. Additionally, we assumed that the spectrum of the signal photon matches the absorption profile of a given sample and that the spectrum of the idler photon allows the use of a free-space single photon detector. Simulation for $\rho = 0$ was run for comparison. In this case, different crystal parameters were chosen such that they led to the same biphoton wavefunction parameters.

\begin{figure}[htb]
	\centering
	\includegraphics[width=0.55\columnwidth]{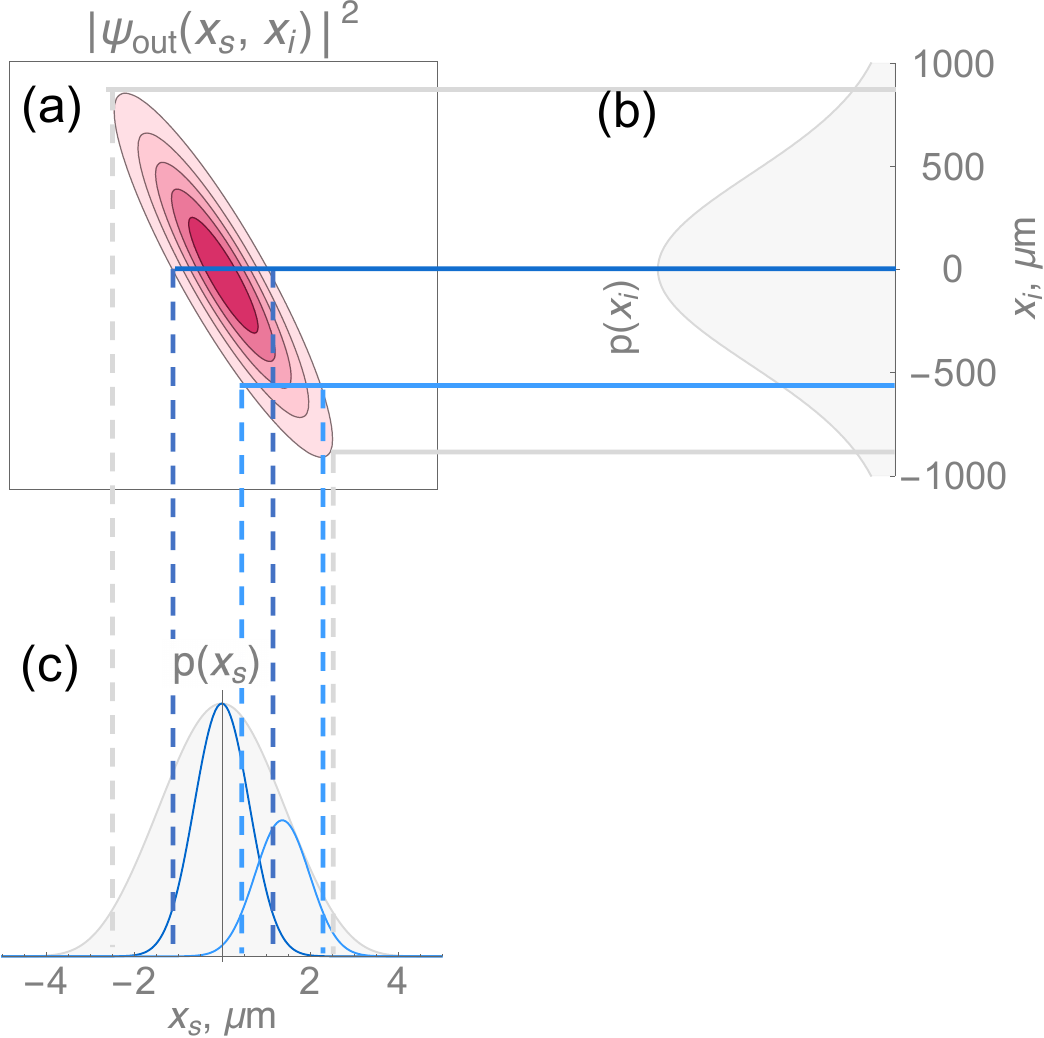}
	\caption{\textbf{Entanglement assisted wavepacket narrowing}   (a) Example joint probability distribution, $|\PSIout(x_\mathrm{s}, x_\mathrm{i})|^2$. The contours are equally spaced. (b) Marginal probability of detecting the idler photon when neglecting information about the signal photon, $p_i(x_\mathrm{i})$ (grey). (c) Marginal probability of detecting the signal photon when neglecting information regarding  idler photon, $p(x_\mathrm{s})$ (grey). Heralded (PLD) signal photon SMPs (blue). The plots were generated for $\rho = 0.9$ and $f_\mathrm{s} = 100$~mm. The horizontal, solid lines at $x_\mathrm{i} = 0$~$\mu$m and $x_\mathrm{i} = -510$~$\mu$m represent the detector's locations.}
	\label{fig:cp125mm4x-cp30mm4x}
\end{figure}

An example simulation result illustrating the effect which is investigated here is given in Fig. \ref{fig:cp125mm4x-cp30mm4x}. Panel (a) presents the joint spatial probability distribution, $|\PSIout(x_\mathrm{s}, x_\mathrm{i})|^2$, computed for correlation coefficient $\rho = 0.9$ and for $f_\mathrm{s} = 100$~mm. The grey Gaussian curves in panels (b) and (c) are the non-heralded probability density distributions, as given in \eqref{eq:p:nh}, of the idler and signal photon detection, respectively. When the idler photon is detected by a PLD at the point in the center, $x_\mathrm{i}=0$~$\mu$m, or at a side $x_\mathrm{i}=-510$~$\mu$m (marked on panel (b) by dark and light blue lines, respectively), then the SMP of the signal photon is given by a function, drawn in the same color on panel (c). FSD results for a commercially available $100$~$\mu$m-active-area detector were obtained as well and are very similar to the PLD ones. 

\begin{figure*}[htb]
	\centering
	\subfigure[]{\includegraphics[width=0.49\linewidth]{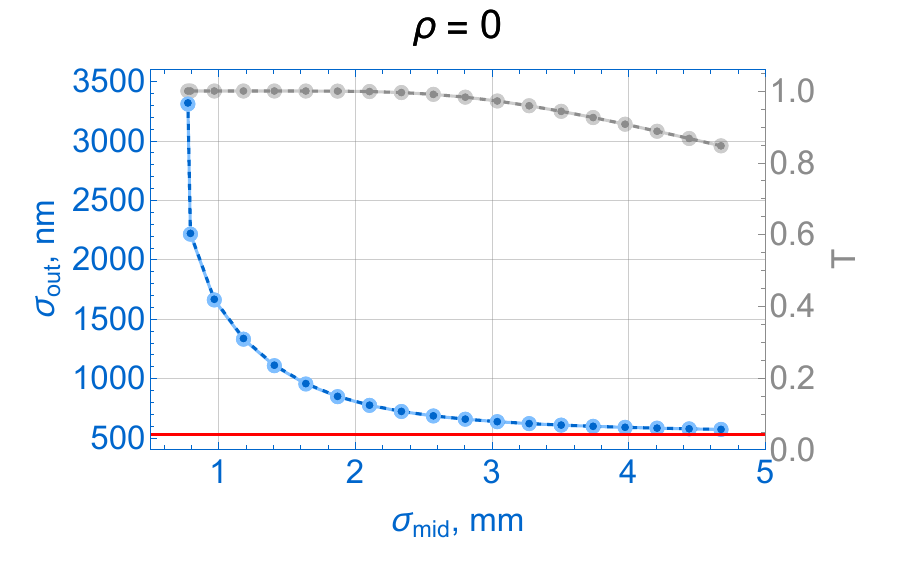}}
	\subfigure[]{\includegraphics[width=0.49\linewidth]{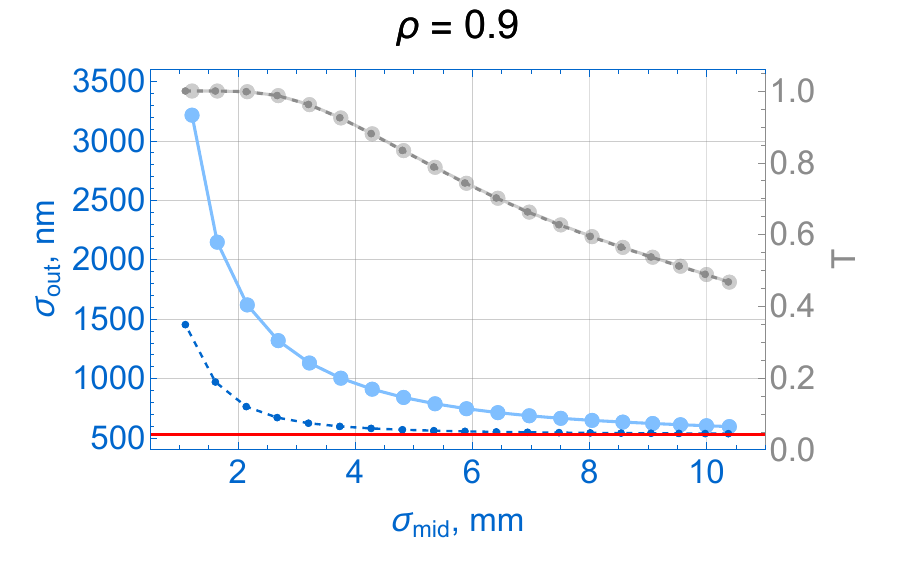}}
	\caption{\textbf{Approaching the diffraction limit.}  Width of the signal photon wavepacket after propagation through the whole optical setup, $\sigma_\mathrm{out}$, together with the transmission of the signal photon, $T$, as functions of the width of the signal photon wavepacket before the MO, $\sigma_\mathrm{mid}$. $\sigma_\mathrm{out}$ is defined as the FWHM of the Gaussian fit to the output wavepacket whereas $\sigma_\mathrm{mid}$ is defined as the radius of an iris able to transmit 99\% intensity of the photon profile described by the wavepacket $\PSImid$. Light-blue, solid: $\sigma_\mathrm{out} (\sigma_\mathrm{mid})$ in the non-heralded scenario. Dark-blue, dashed: $\sigma_\mathrm{out} (\sigma_\mathrm{mid})$ in the heralded scenario with PLD placed at $x_i=0$. Light-grey, solid: $T (\sigma_\mathrm{mid})$ in the non-heralded scenario. Dark-grey, dashed: $T (\sigma_\mathrm{mid})$ in the heralded scenario. The diffraction limit is marked as a red solid line which was obtained by propagating a delta function, $\delta(x) $, through the optical setup and from a Gaussian fit the width of $\sigma_{out, \delta(x)} = \sqrt{2 ln 2}  \cdot \sigma_{\delta(x)} =  526.22$~nm was obtained, where $\sigma_{\delta(x)} $ is the Gaussian function fit parameter. The simulation was run for $f_\mathrm{s} \in (100, 1000)$~mm and $\rho = 0, 0.9$, where $f_\mathrm{s}$ was the parameter used for the signal photon SMP before the MO, $\sigma_\mathrm{mid}$, determination. $f_\mathrm{s} = 100$ mm corresponds to the smallest $\sigma_\mathrm{mid}$ and $f_\mathrm{s} = 1000$ mm to the biggest. Note, that the $\sigma_\mathrm{mid}$ takes different values for the same $f_\mathrm{s}$ in the case of different correlations $\rho$ and hence leads to different $\sigma_\mathrm{mid}$-axis ranges.}
	\label{fig:overlayTEST-0-9-99-t99}
\end{figure*}

The analysis of the signal photon SMP in the sample plane together with the transmission through the MO for a complete set of parameters is given in \figref{fig:overlayTEST-0-9-99-t99}. The \textit{x-axis} of each plot represents the width of the SMP, $\sigma_\mathrm{mid}$, at the entrance to the MO. Here, the width, $\sigma_\mathrm{mid}$, is defined as the radius of an iris, which is able to transmit 99\% of the photons. The left \textit{y-axis} represents the FWHM of the Gaussian fit to the output SMP at the sample plane, $\sigma_\mathrm{out}$, whereas the right \textit{y-axis} stands for the wavepacket transmission through the MO. Four figures of merit are presented on each plot as a function of the SMP's width before the MO: 1) output SMP width in the non-heralding scenario (light blue, dotted), 2) output SMP width in the heralding, PLD scenario  (dark blue, dotted-dashed), 3) transmission of the wavepacket through the MO in the non-heralding scenario (light-grey, dashed), and 4) transmission of the wavepacket through the MO in the heralding scenario (dark-grey, dashed). In the heralding scenario the PLD was placed at $x_i=0$ but placing the detector at any other position leads to the same result, as our numerical simulation results confirm and our analytical study supports.

First, let's discuss the SMP characteristics at the sample plane in the heralding and non-heralding scenarios. In the case where the photon pair features no entanglement, $\rho = 0$, the width, $\sigma_\mathrm{out}$ is identical in both,  heralding and non-heralding, scenarios as it can be seen in \figref{fig:overlayTEST-0-9-99-t99}(a). Moreover, as expected, a bigger spatial mode entering the MO corresponds to a smaller SMP in the sample plane. \figref{fig:overlayTEST-0-9-99-t99}(b) displays the difference when entanglement is introduced. The monotonicity of the $\sigma_\mathrm{out}$ as a function of $\sigma_\mathrm{mid}$ is preserved, but the sets of points for the heralding and non-heralding scenarios do not overlap anymore. It can be seen that for a given width of the spatial mode entering the MO, $\sigma_\mathrm{mid}$, one cannot achieve in the non-zero correlations case and non-heralding scenario as small spot size as in the zero-correlations one. To approach the diffraction limit, heralding is necessary. For instance, $\rho=0.9$ for $f_\mathrm{s} = 200$~mm, which corresponds to around $\sigma_{\text{mid}}= 2.15$ mm, leads to the resolution of approximately $1.62~\mu$m in the non-heralding scenario, whereas $0.76~\mu$m resolution is obtained when heralding. This result is interesting from the fundamental point of view. It shows that when using SPDC source to illuminate a sample with quantum light and therefore dealing with spatially correlated photon pairs it is not sufficient to have the spatial mode of a comparable size to the focusing lens aperture. In this case, a heralding scenario has to be additionally applied to approach the diffraction limit.

When working with single photons it is very important to analyze  the diffraction limit together with the transmission through the MO. What can be observed in  \figref{fig:overlayTEST-0-9-99-t99}(b) is that in the case of non-zero correlations and a non-heralding scenario (light-blue, solid), the diffraction limit can be approached with the transmission below $0.7$. However, when the signal photon is heralded at $x_i=0$ (dark-blue, dashed), the diffraction limit with a much greater transmission, above $0.95$, is achieved. Nevertheless, the heralded single-photon transmission and heralding efficiency both decrease when heralding at $x_i\neq 0$. This is related to the heralded mode central position tracking according to \eqref{eq:EAST} and the heralding efficiency proportionality relation to the non-heralded mode intensity profile. These are important observations in the context of applications making use of correlated photon pairs, like quantum ghost imaging or quantum microscopy, where any loss is detrimental. 

\noindent

Analogous results to those presented in \figref{fig:cp125mm4x-cp30mm4x} and \figref{fig:overlayTEST-0-9-99-t99} were obtained in a scenario were a PLD was replaced with a finite size detector (FSD). A commercially available, $100$~$\mu$m-active-area FSD was chosen and it resulted in a negligible difference in the SMP of the signal photon. It was not bigger than $1$\% for all sets of parameters. Also, we determined the maximal active area of the FSD, which still approximates the PLD well. This was done by placing a heralding PLD detector in the signal photon arm and investigating the SMP of the idler photon in a heralding scheme designed this way. The SMP of the heralded idler photon settled the maximal size of the FSD for a given set of parameters. For the $L_s$ lens focal length equal to the extreme values of the simulated range, it resulted in the FSD active area on the order of $480$~$\mu$m for $f_s = 100$~mm and $960$~$\mu$m for $f_s=1000$~mm. We found out that such replacement would correspond to respectively $4$\% and $18$\% change in the SMP width of the signal photon.

\section{Summary and outlook}

Summarizing, the SMP of a heralded single photon in a quantum microscopy setting was analyzed. Analytical formulas for the  infinite-size optics scenario were provided and numerical calculations were presented for the finite-size optics one. The analysis was done for an experimentally accessible set of parameters including finite-size detectors.
It was demonstrated that the microscope resolution depends not only on the spatial mode of the photon entering the MO but also on its phase profile. This was shown to be accurate both for quantum and classical light. We have provided analytical formulas on how to prepare a phase-altering device at the entrance to a lens to manipulate the SMP of a Gaussian beam propagated through the lens. In particular, when non-optimal distance between signal arm lenses was chosen, and hence a non-zero quadratic phase before MO appeared causing the heralded SMP broadening, it was shown that it could be compensated by an adaptive optics device. 

Additionally, the presented work constituted a framework for a quantum light illumination in a ghost-imaging microscopy setting. Multi-photon Fock states were already proven to enable beating the diffraction limit \cite{Simon2016, Boto2000} and hence resulted in the lateral resolution improvement. Similarly, the correlation in the number of photons may enable resolution improvement.

%\appendix
%\appendixpag
\begin{appendices}
\section{Biphoton wavefunction propagation}
\label{app:derivation}

\renewcommand{\theequation}{A-\arabic{equation}}
\setcounter{equation}{0}  % reset counter 

To analyze the signal photon SMP in a chosen spot of the experimental setup depicted in \figref{Fig:Experimental-setup-concept}, the biphoton wavefunction \eqref{eq:wavefunction2} has to be propagated. In the main text the signal photon SMP is analyzed in two characteristic spots: just before the MO and in the focal plane of the MO, which is the sample plane. This is done by evaluating analytically \eqref{eq:wavefinction-after-propagation-general-idea} with appropriate propagators. Now, the propagation up to the two aforementioned characteristic spots will be analyzed separately.

\subsection{Before the MO}

To obtain the analytical formula for the signal photon SMP at the entrance to the MO \eqref{eq:wavefinction-after-propagation-general-idea} has to be analytically evaluated using a propagator related to the full path propagation in the idler arm and propagation
up to the MO in the signal arm. This corresponds to

\begin{multline}
	\psi_{mid} (x_s', x_i) = \int_{-\infty}^{+\infty} dx_i' S_x(x_i',x_i,d_{i3}) L(x_i', f_{i2}) \\
	\times \int_{-\infty}^{+\infty} dx_s'' \int_{-\infty}^{+\infty}  dx_i'' S_x (x_s'', x_s', d_{s2}) S_x (x_i'', x_i', d_{i2}) L(x_s'', f_s) L(x_i'', f_{i1})\\ 
	\times \int_{-\infty}^{+\infty} dx_s''' \int_{-\infty}^{+\infty}  dx_i''' S_x (x_s''', x_s'', d_{s1}) S_x (x_i''', x_i'', d_{i1}) \times \psi_{in} (x_s''', x_i'''),
	\label{eq:psi_mid_derivation_1}
\end{multline}
where $S_x (x', x, d) = \exp\left( -{i \pi (x' - x)^2}/{\lambda d}\right)$ stands for the free-space propagator over a given distance $d$ and $L(x, f) = \exp\left( {i \pi x^2}/{\lambda f}\right)$ stands for the lens of focal length $f$ propagator. This, in turn, can be formulated as

\begin{multline}
	\psi_{mid} (x_s', x_i) = \int_{-\infty}^{+\infty} dx_s''' \int_{-\infty}^{+\infty} dx_i''' \psi_{in} (x_s''', x_i''') \\
	\times P_{mid} (x_s''', x_s', x_i''', x_i, f_s, f_{i1}, f_{i2}, d_{s1}, d_{s2}, d_{i1}, d_{i2}, d_{i3}),
	\label{eq:psi_mid_derivation_2}
\end{multline}
where

\begin{multline}
	P_{mid} (x_s''', x_s', x_i''', x_i, f_s, f_{i1}, f_{i2}, d_{s1}, d_{s2}, d_{i1}, d_{i2}, d_{i3}) = \\
	= \int_{-\infty}^{+\infty} dx_i' \int_{-\infty}^{+\infty} dx_s'' \int_{-\infty}^{+\infty} dx_i'' S_x(x_i',x_i,d_{i3}) L(x_i', f_{i2})  S_x (x_s'', x_s', d_{s2}) S_x (x_i'', x_i', d_{i2}) \\
	\times L(x_s'', f_s) L(x_i'', f_{i1}) S_x (x_s''', x_s'', d_{s1}) S_x (x_i''', x_i'', d_{i1})
\end{multline}
is the effective propagator in \eqref{eq:wavefinction-after-propagation-general-idea}.

The input function, $\psi_{in}$, and the applied propagators are of a Gaussian form. Hence, the integrands in \eqref{eq:psi_mid_derivation_1} and \eqref{eq:psi_mid_derivation_2} are also of a Gaussian form. It can be easily checked that integration of a Gaussian function over infinite limits as a result gives again a Gaussian function. The probability density of the non-heralded signal photon detection at the entrance to the MO can be calculated as $\int_{-\infty}^{\infty} dx_\mathrm{i} |\PSImid(x_\mathrm{s}', x_\mathrm{i})|^2$ and hence the result will be of a general form

\begin{equation}
	|\psi_1(x_s)|^2 = |N_1|^2 \exp{\left(- 2\frac{(x_s-x_{s1})^2}{\sigma_1^2}\right)},
	\label{eq:psi1_squared}
\end{equation}
where $\psi_1(x_s)$ is a 1-dimensional Gaussian function, given by \eqref{eq:single_beam_profile} in the main text. In this way, the $\sigma_1$ parameter can be found which in this case corresponds to the width of the non-heralded signal photon SMP before the MO expressed explicitly by \eqref{eq:width_before_MO_non_heralded} in the main text.

On the other hand, the probability density of the heralded signal photon detection at the entrance to the MO can be calculated as $|\PSImid(x_\mathrm{s}', x_\mathrm{i})|^2$. The result will be again of the form given by \eqref{eq:psi1_squared}. This time the $\sigma_1$ parameter would correspond to the width of the heralded signal photon SMP before the MO given in \eqref{eq:width_before_MO_heralded} in the main text.

The formulas given in \eqref{eq:A} -- (\ref{eq:C}) are derived similarly by noticing that the imaginary part of the index of the $\PSImid(x_\mathrm{s}', x_\mathrm{i})$ function is of a general form given by \eqref{eq:phase}.

\subsection{Sample plane}

To obtain the analytical formula for the signal photon SMP at the sample plane \eqref{eq:wavefinction-after-propagation-general-idea} has to be analytically evaluated using a propagator related to the full path propagation in both the idler and signal photon arms. This corresponds to

\begin{multline}
	\psi_{out} (x_s', x_i) = \int_{-\infty}^{+\infty} dx_s' \int_{-\infty}^{+\infty} dx_i' S_x(x_s',x_s,d_{s3}) S_x(x_i',x_i,d_{i3}) L(x_s', f_{MO}) L(x_i', f_{i2}) \\
	\times \int_{-\infty}^{+\infty} dx_s'' \int_{-\infty}^{+\infty}  dx_i'' S_x (x_s'', x_s', d_{s2}) S_x (x_i'', x_i', d_{i2}) L(x_s'', f_s) L(x_i'', f_{i1})\\ 
	\times \int_{-\infty}^{+\infty} dx_s''' \int_{-\infty}^{+\infty}  dx_i''' S_x (x_s''', x_s'', d_{s1}) S_x (x_i''', x_i'', d_{i1}) \times \psi_{in} (x_s''', x_i'''),
	\label{eq:psi_full_derivation_1}
\end{multline}
After some algebra, this can be formulated as

\begin{multline}
	\psi_{out} (x_s', x_i) = \int_{-\infty}^{+\infty} dx_s''' \int_{-\infty}^{+\infty} dx_i''' \psi_{in} (x_s''', x_i''') \\
	\times P_{out} (x_s''', x_s, x_i''', x_i, f_s, f_{MO}, f_{i1}, f_{i2}, d_{s1}, d_{s2}, d_{s3}, d_{i1}, d_{i2}, d_{i3}),
	\label{eq:psi_full_derivation_2}
\end{multline}
where

\begin{multline}
	P_{out} (x_s''', x_s, x_i''', x_i, f_s, f_{MO}, f_{i1}, f_{i2}, d_{s1}, d_{s2}, d_{s3}, d_{i1}, d_{i2}, d_{i3}) = \\
	= \int_{-\infty}^{+\infty} dx_s' \int_{-\infty}^{+\infty} dx_i' \int_{-\infty}^{+\infty} dx_s'' \int_{-\infty}^{+\infty} dx_i'' S_x(x_s',x_s,d_{s3}) S_x(x_i',x_i,d_{i3}) L(x_s', f_{MO}) L(x_i', f_{i2}) \\
	\times S_x (x_s'', x_s', d_{s2}) S_x (x_i'', x_i', d_{i2}) L(x_s'', f_s) L(x_i'', f_{i1}) S_x (x_s''', x_s'', d_{s1}) S_x (x_i''', x_i'', d_{i1})
\end{multline}
is the effective propagator in \eqref{eq:wavefinction-after-propagation-general-idea}.

Similarly as in the previous subsection, the input function, $\psi_{in}$, and the applied propagators are of a Gaussian form. Hence, the integrands in \eqref{eq:psi_full_derivation_1} and \eqref{eq:psi_full_derivation_2} are also of a Gaussian form. The probability density of the non-heralded signal photon detection at the sample plane can be calculated as $\int_{-\infty}^{\infty} dx_\mathrm{i} |\PSIout(x_\mathrm{s}', x_\mathrm{i})|^2$. The result will be again of the form given by \eqref{eq:psi1_squared}.  In this way, the $\sigma_1$ parameter can be found which in this case corresponds to the width of the non-heralded signal photon SMP at the sample plane given explicitly by \eqref{eq:non-herladed-width-after-MO} in the main text.

On the other hand, the probability density of the heralded signal photon detection at the sample plane can be calculated as $|\PSIout(x_\mathrm{s}', x_\mathrm{i})|^2$. The result will be again of the form given by \eqref{eq:psi1_squared}. This time the $\sigma_1$ parameter would correspond to the width of the heralded signal photon SMP at the sample plane given in \eqref{eq:herladed-width-after-MO} in the main text.

\end{appendices}

\section*{Funding}
HORIZON EUROPE Framework Programme (101070062); Brandon University; Narodowe Centrum Nauki (2019/35/N/ST2/03443).

\section*{Acknowledgments}
All the authors acknowledge support by the National Laboratory of Atomic, Molecular and Optical Physics, Toru\'{n}, Poland. MG and PK are grateful for insightful discussions with \mbox{Sylwia M. Kolenderska}.

\section*{Disclosures}
The authors declare no conflicts of interest.

\section*{Data availability} No data were generated or analyzed in the presented research.

\bibliography{maria}

\end{document}